# New insights on the formation of supersaturated solid solutions in Cu-Cr system deformed by High-Pressure Torsion

A. Bachmaier[a*], G.B. Rathmayr[b], M. Bartosik[cI], D. Apel[d], Z. Zhang[b], R. Pippan[b]

[a]Chair of Materials Science and Methods, Saarland University, Campus D2.2, Saarbrücken, 66123, Germany
[b]Erich Schmid Institute of Materials Science, Austrian Academy of Science, Leoben, 8700, Austria
[c]Department Materials Physics, University of Leoben, Leoben, 8700, Austria
[d]Helmholtz-Zentrum Berlin für Materialien und Energie, Albert-Einstein-Straße 15, 12489 Berlin, Germany
[I]Present address: Institute of Materials Science and Technology, Vienna University of Technology, Karlsplatz 13/E308, Vienna, 1010, Austria
*Corresponding author: a.bachmaier@matsci.uni-sb.de

**Abstract**

In the Cu-Cr system, the formation of supersaturated solid solutions can be obtained by severe plastic deformation. Energy-dispersive synchrotron diffraction measurements on as-deformed Cu-Cr samples as a function of the applied strain during deformation confirm the formation of supersaturated solid solutions in this usually immiscible system. Due to evaluation of the diffraction data by a newly developed energy-dispersive RIETVELD-program, lattice parameter and microstructural parameters like domain size and microstrain are determined for as-deformed as well as annealed samples. The obtained information is used to deepen the understanding of the microstructural evolution and the formation of supersaturated solid solutions during severe plastic deformation. Complimentary transmission electron microscopy investigations are furthermore performed to characterize the evolving microstructure in detail. After annealing at elevated temperatures, the formed solid solutions decompose. Compared to the as-deformed state, an enhanced hardness combined with a high thermal stability is observed. Possible mechanisms for the enhanced hardness are discussed.

# 1. Introduction

Ultrafine grained materials, as well as nanocrystalline (nc) materials, can be produced by severe plastic deformation (SPD) from nearly all kinds of bulk coarse grained metal materials [1,2]. In the last few years, the effect of SPD on composites or multiphase materials, which can lead to nanocomposites with a grain size in the order of about 10 nm, has also been studied [3]. During SPD of composites or powder mixtures, the formation of supersaturated solid solutions (ssss) or even amorphization reactions can occur in alloys with a positive heat of mixing ΔH, due to a kind of "mechanically alloying in bulk form" [4-13].

The crystalline structure of Cu is face-centered cubic (fcc), Cr exhibits a body-centered cubic (bcc) structure. This binary system exhibits a positive Gibbs free energy change $\Delta G_c$ for forming a solid solution alloy in the equilibrium state and is therefore nearly immiscible in the solid state [14]. However, the formation of ssss in the Cu-Cr system during ball milling (BM) has been observed [15-17].

SPD induced solid state amorphization is not observed in the Cu-Cr system which is frequently observed in immiscible systems which partly fulfill glass-forming rules [18, 19]. In [20], a thermodynamic analysis was conducted to predict the stable phases in the Cu-Cr system for the whole composition range. $\Delta G_c$ for the formation of a crystalline solid solution is lower than for an amorphous phase but both are positive, i.e. ssss are more stable than the amorphous phase.

In [21], the formation of ssss in a Cu-Cr composite material subjected to SPD by means of High-Pressure Torsion (HPT) is reported as well. Certain differences between BM and HPT exist: During HPT the strain rate is constant and very small. During BM, the powders experience high strain rate plastic deformation due to the impact of the milling balls. Since the strain rate is small, the temperature during HPT deformation is not significantly rising and nearly constant. Contrary, the local temperature can rise due to the impact of the milling balls during BM. Furthermore, there is no contamination in case of HPT [22]. Therefore, ssss formation seems to be only an effect of heavy plastic deformation and not of the mentioned special phenomena occurring during BM. This finding is furthermore supported by the fact that ssss formation observed during SPD are also independent of the used SPD method [4-13].

The strain applied during HPT deformation leads to a significant refinement of the initially coarse grained Cu-Cr material with a final grain size of 10–20 nm as shown by transmission electron microscopy (TEM) investigations [21]. Both, the fcc Cu phase and the bcc Cr phase are present in the nanostructure. Atom probe tomography (APT) investigations have proven that a significant amount of Cu is dissolved in the Cr phase in the as-deformed condition. During annealing at 450°C for 30 minutes, phase separation and grain growth occurred leading to a composite material consisting of Cr grains free of Cu with a grain size of 20–40 nm with an enhanced hardness compared to the as-deformed condition. The details of these phenomena however remains unclear.

In literature, the physical mechanisms behind mechanical alloying are still extensively discussed. Thermodynamic considerations include that the positive heat of mixing can be overcome by the high interfacial energy arising from the nanometer sized phases together with the high number of defects introduced during the deformation [23-26]. Mechanical intermixing mechanisms, which are driven by the plastic deformation involving trans-phase dislocation shuffling are also proposed in literature [18,19]. Dislocation transfer from one to the other phase is also important in the shear induced chemical mixing model presented in [27,28].

The present work is again focused on HPT deformed and annealed Cu-Cr composite material. The aim is to gain a deeper insight of the evolution of the microstructure and the deformation-induced formation of ssss in the Cu-Cr composite during HPT processing and decomposition during subsequent annealing treatments. Energy-dispersive synchrotron diffraction measurements as a function of the equivalent strain ($\varepsilon_{eq}$) of the deformed and annealed samples are performed. Energy-dispersive synchrotron diffraction provides fast measurement of complete diffraction patterns with a multitude of diffraction lines under fixed but arbitrary scattering directions. RIETVELD refinement of the energy-dispersive diffraction (EDD) patterns enables particle size and strain broadening analysis of the Cu-Cr material [29]. The EDD patterns contain a variety of information, e.g. the lattice parameter and structural material properties like microstrain and domain size of both phases in the composite, which provide new insight of the microstructural evolution during HPT deformation, formation of ssss and decomposition during annealing. The whole EDD patterns are analysed using a recently developed energy-dispersive RIETVELD method [29]. The information obtained from EDD measurements is combined with complementary hardness measurements and TEM investigations.

## 2. Experimental

A coarse-grained Cu-Cr composite material (43wt. % Cr, 57wt. % Cu) was HPT deformed at room temperature (RT). The initial microstructure of the composite is shown in [21], which consists of a Cu matrix with Cr particles (volume fraction of about 50%, mean diameter of about 50 µm) produced by PLANSEE (Reutte, Austria). Disks with a diameter of 8 mm and a thickness (*t*) of about 0.8 mm were HPT deformed for different numbers of turns *n* (*n*=1, 5, 25, 100) under a constant pressure of 6.25 GPa with a rotation speed of 0.2 rotations/minute. HPT disks which were deformed for 25 and 100 turns were additionally annealed at temperatures of 100°C, 200°C, 300°C, 400°C, 550°C and 650°C for 30 minutes after the deformation. All data shown in this paper is either presented as a function of $\varepsilon_{eq}$ or given for a certain $\varepsilon_{eq}$ (radius *r*=3.5 mm from the disk centre), respectively. $\varepsilon_{eq}$ is calculated by [1,2]:

$$\varepsilon_{eq} = \frac{2 \cdot \pi \cdot r \cdot n}{t \cdot \sqrt{3}} \ . \tag{1}$$

Vickers microhardness measurements were conducted on a BUEHLER Mircomet 5100 using a load of 500 g ($HV_{0.5}$). Indents were made across the radii of the disks with a spacing of 250 µm between the indents and mean values of at least three separate measurements on three different HPT deformed disks for each deformation condition and each position are reported in this work. Microhardness measurements were further conducted on the annealed disks across the radii of the disks at each annealing condition as described above.

TEM and scanning TEM (STEM) was used to characterize the microstructure of the deformed material which was deformed for 25 rotations in detail. All microstructural investigations were undertaken at r=3.5 mm from the torsion axis of the HPT disks which corresponds to εeq of about 400. After an additional annealing treatment (400°C, 30 min), the microstructure was investigated at the same position. The TEM sample preparation steps are as follows: Disks were cut from the HPT samples, which were mechanically thinned and polished to a thickness of about 100 μm, followed by mechanical dimpling until the thinnest part reached a thickness of about 10 μm. Subsequently the samples were ion-milled with Ar ions at 5 kV under an incidence angle of 5 ° using a Gatan Precision Ion Polishing System until perforation was obtained. In a final step, ion polishing with low-energy Ar ions was performed at 1.5 kV under an incidence angle of 4 ° for 20 minutes. TEM studies were carried out using a field emission gun transmission electron microscope (JEOL JEM-2100F) equipped with an imaging spherical aberration corrector and an Oxford INCA Energy TEM 200 energy-dispersive X-ray spectroscopy (EDS) system. The electron beam was perpendicular to the shear plane of the disks for all microstructural investigations shown in this work. The STEM-HAADF detector was used with a spot size of 1.0 nm and a converge angle of 15 mrad. While recording STEM HAADF image, the STEM detector inner angle and outer angle used were 86 mrad and 230 mrad, respectively. Under this condition, the STEM HAADF image is with Z- contrast. EDS for nanoscale compositional analysis were also carried out in STEM mode. The chemical composition was calculated by standard loss method.

Synchrotron X-ray experiments were performed at the materials science beamline for *E*nergy *D*ispersive *DI*ffraction (EDDI) at the synchrotron facilities BESSY II (Berlin) [30]. The EDD measurements were carried out on deformed (*n*=5, 25, 100 revolutions) and annealed disks (*n*=25+annealing at 200°C, 300°C, 400°C) at different positions across the radii of the disks next to the indents of the hardness measurements with the same spacing of 250 µm between each measurement. The primary beam cross section was set to 0.5 x 0.5 mm². On the secondary side the apertures of the two-slit system were adjusted to 0.03 x 5 mm² (equatorial x axial) to avoid geometrically induced line-broadening and to restrict line-shifts to < 0.01°. The counting time for each EDD pattern was 30s using a diffraction angle *2θ*=12°. The evaluation of the EDD data as function of $\varepsilon_{eq}$ was conducted using a recently developed energy-dispersive RIETVELD-program to determine the lattice parameter and microstructural parameters like domain size and microstrain, for details see [29]. The RIETVELD-program makes use of a modified Thomson, Cox and Hastings [31] line profile model to determine the microstructural parameters. For the calculation of the structure factors the LE BAIL method [32] was used. To account for the instrumental broadening the refined values of the microstructural parameters of the sample under investigation were corrected for instrumental effects, which were determined using the LaB6 SRM660b (NIST) powder. For the two-phase system, RIETVELD refinement of the refinable parameters were done separately for each phase using five peaks for the Cu phase (111, 200, 220, 311, 222) and four peaks for the Cr phase (110, 200, 211, 220). During the refinement of the Cu phase a larger residual for the 200 peak was observed (Fig.1). Neither the peak position nor the shape could be described adequately using the line profile model offered by the RIETVELD-program used for the analysis. A possible explanation for this observation could be the occurrence of anisotropic plastic deformation of the Cu phase during the sample treatment resulting in anisotropic lattice parameter changes and anisotropic microstructural properties. A model to describe such anisotropic material properties is not yet implemented in the RIETVELD-program. Yet, to be able to investigate the lattice parameter and the microstructural properties of the Cu phase, the following approach was carried out. The peaks of the Cu phase were divided into two groups: group one contains the 111 and 222 peak (both peaks have the same crystallographic direction and the same diffraction elastic constants (DEC)) and group two contains the 200, 220 and 311 peak (those peaks have similar crystallographic directions and have similar DEC). The RIETVELD refinement was then done separately for each group. For group one (111 and 222 peak) the RIETVELD refinement delivers reliable results. The refinement for the second group (200, 220, 311 peak) was performed for completeness and resulted in less pronounced residuals. In this work, mainly the results of group one are considered and discussed.

3. Results

The microhardness for Cu-Cr samples as a function of the applied strain $\varepsilon_{eq}$ is shown in Fig.2. The hardness of the undeformed Cu-Cr sample is ~100 HV, which increases rapidly to ~200 HV in the beginning of the deformation. Only small changes of the hardness are observable until a strain $\varepsilon_{eq}$ of 50. Subsequently, a strong hardness increase to ~400 HV until an applied strain $\varepsilon_{eq}$ of 200, followed by a slight further increase to 450 HV at the highest deformation degree, is visible. Only a small increase of the hardness values are observed although a three time higher strain is applied between a $\varepsilon_{eq}$ of 500 to 1600. Hardness values of this study fit quite well to hardness data already reported in [21]. The EDD patterns for the sample deformed for 25 rotations recorded at

r=0 mm ($\varepsilon_{eq}$ ~0) and r=3.5 mm ($\varepsilon_{eq}$ ~400) is shown in Fig.3a. In the "undeformed" state, peaks of the fcc Cu phase and the bcc Cr phase are visible. Both sets of diffraction peaks are well-defined and narrow. In the EDD pattern of the deformed state ($\varepsilon_{eq}$ ~400), both sets of peaks are still visible, but significantly broadened, and partially overlapped with each other. The intensity of the diffraction peaks is decreased, whereby the intensity of the Cu diffraction peaks reduces stronger compared to the Cr peaks. The peaks of the Cr phase are also shifted to lower energies in the deformed state compared to those recorded at the centre of the disk, i.e. in the undeformed state. In Fig.3b-d, the calculated values for domain size, lattice parameter and microstrain of the Cu-Cr sample deformed for 25 rotations as a function of $\varepsilon_{eq}$ for the Cr and Cu phase are shown. The calculated values for the Cu (111, 222) and Cu (200, 220, 311) peaks are plotted separately in the graphs. The domain size of the Cr phase decreases from ~38 nm in the centre to ~8 nm at the outer edge of the disk. Above an applied strain $\varepsilon_{eq}$ of 200, no much additional microstructural refinement occurs and the domain size reduces only slightly from 12 to 8 nm with increasing strain. The domain size of the Cu phase is decreasing as well. The domain size, which is calculated by the Cu (111, 222) peaks, reduces from ~48 nm in the centre to ~20 nm at the outer edge of the disk. In contrast, a far smaller domain size is calculated for the Cu (200, 220, 311) peaks as a function of $\varepsilon_{eq}$. The domain size in this case is similar to the one of the Cr phase. The different domain size which is determined for the (111, 222) and (200, 220, 311) peaks of the Cu phase might be due to an elongated shape of the Cu grains or to twins and/or stacking faults which lead to an apparent particle size smaller than the true particle size [33]. Calculated values for domain size, lattice parameter and microstrain for three different deformation conditions are listed in Table 1. Even if a large additional strain is applied, the domain size of the Cr and Cu phase is nearly unchanged and no further grain refinement takes place even at a $\varepsilon_{eq}$ of 1590. The minimum domain size of Cr and Cu in the Cu-Cr composite material is achieved. In Fig.3c, the change of the lattice parameter is plotted. The lattice parameter of the Cr phase increases with increasing deformation, i.e. increasing $\varepsilon_{eq}$. The lattice parameter change for Cu, calculated by the (200, 220, 311) peaks, is quite similar. The lattice parameter for Cu, calculated by the (111, 222) peaks, stays nearly constant. In Fig. 3d, the calculated microstrain as a function of $\varepsilon_{eq}$ is plotted for each phase. The microstrain of the Cr phase increases slightly with increasing deformation at low strains. Between a $\varepsilon_{eq}$ of 100-150, a constant value of about $0.3 \cdot 10^{-2}$ is reached. For $\varepsilon_{eq}$ larger than 175, the calculated values of the microstrain for the Cr phase monotonously reduce to values near zero. At high strains, either no or only a small amount of microstrain, non-detectable with the instrument, can be measured. At low strains, the microstrain of the Cu phase is quite similar or even less to the values calculated for Cr. Values calculated for the (200, 220, 311) peaks of the Cu phase are somewhat larger than those calculated for the (111, 222) peaks. At a $\varepsilon_{eq}$ larger than 150, the microstrain is steadily increasing and a high microstrain of $1.5 \cdot 10^{-2}$ is measured at the highest degree of deformation (see Table 1). For the Cr phase, only a microstrain of $0.3 \cdot 10^{-2}$ is calculated at the same $\varepsilon_{eq}$. The observed broadening in the EDD pattern in the as-deformed condition is mainly due to the refinement of the grain (domain) size during SPD. The considerable broadening in the case of the Cu phase is additionally caused by the microstrain.

In Fig.4a, STEM HAADF images with Z-contrast of the Cu-Cr sample after HPT processing to a $\varepsilon_{eq}$ of 400 are shown. The composite consists of nc Cr and Cu grains with a size below 20 nm. The Cu phase seems to be slightly elongated in the as-deformed state. Many grain boundaries are not distinct, but appear to be curved. EDS analysis was carried out to determine the local chemical composition along lines across interphase boundaries in STEM mode. The chemical composition and distribution of Cu and Cr elements in atomic percent (at. %) on the nano-scale level are shown in the concentrations profiles plotted in Fig.4c and d (line 1 and 2 marked in the STEM images). For the concentration profile shown in Fig.4c and Fig.4d, 10 measurements with 3 nm spacing between a single measurement and 20 measurements with 1 nm spacing were made, respectively. The intensity profiles of the characteristic X-ray spectral lines demonstrate that the composition varied with the position across the interface regions along the lines. Along line 1, a Cr phase free of Cu, a Cu phase free of Cr and a Cr phase containing ~20at.% Cu is visible. Along line 2, a Cr phase in which ~25at.% Cu is dissolved can be seen. The data suggests, that ssss based on Cr is formed during HPT processing, whereas nearly no or only a low content Cr is dissolved in the Cu matrix.

Fig.5a illustrates the EDD patterns recorded at a radius of 3.5 mm ($\varepsilon_{eq}$~400) for samples deformed for 25 rotations with subsequent annealing treatments at 200°C, 300°C and 400°C for 30 min after the deformation. It can be seen that in all three annealing conditions, peaks of the fcc Cu phase and the bcc Cr phase are visible. No significant effect of the annealing on peak broadening is visible for annealing temperatures of 200°C and 300°C. The peaks are still broadened and partly overlap with each other. In the Cr phase, the broadening is slightly decreasing with increasing annealing temperature. An increase of the temperature to 400°C results in a sharpening of the Cu and Cr peaks after the annealing. In Fig.5b-d and Table 1, the calculated values for domain size, lattice parameter and microstrain for the Cr and Cu phase in the as-deformed state at $\varepsilon_{eq}$ of 0 and 400 and as a function of the annealing temperature (200°C, 300°C, 400°C) are shown. The domain size of the Cr phase is nearly unchanged after annealing at 200°C and 300°C and slightly increases at 400°C. The domain size of the Cu phase behaves differently. Calculated by (200, 220, 311) peaks, it increases at 200°C and 300°C, whereas calculated by (111, 222) peaks, it decreases. At 400°C, values of 18 nm and 9 nm for the Cu phase are determined. Discrepancies of the domain size during annealing might be related to twins and/or stacking faults which lead to an apparent particle size smaller than the true particle size [33]. The lattice parameter of both phases change only slightly after annealing at 200°C and 300°C. At 400°C, the initial values of the "undeformed" state for both phases are almost reached. In Fig.5d, the calculated microstrain is plotted. The microstrain of the Cu phase calculated by the (200, 220, 311) peaks increases slightly further at annealing temperatures of 200°C. Values calculated for the (111, 222) peaks are still smaller than those calculated for the (200, 220, 311) peaks. In the Cr phase, either no or only a small amount of microstrain, non-detectable with the instrument, can be measured in the annealed samples. After annealing at 400°C, only a small amount of microstrain ($0.06 \cdot 10^{-2}$) in the Cu phase can be found, too. It can be seen, that the microstrain follows the change of the lattice parameters in both phases during annealing at 200°C and 300°C similar to the as-deformed state [34].

In Fig.6, STEM HAADF images with Z-contrast with high and low magnification of the Cu-Cr sample after HPT processing ($\varepsilon_{eq}$~400) with additional annealing for 30 min at 400°C are shown. After the annealing treatment, the grain size in the composite is larger compared to the as-deformed condition, but still nc. The grain size range in the annealed state (20-40 nm) is comparable to [21]. Twins can also be seen in some of the grains in the STEM image at low magnification. The nc structure is more equiaxed and the Cu and Cr interphase boundaries are distinct. EDS analysis was carried out and the distribution of Cu and Cr after annealing are shown in the concentration profiles (Fig.6c and d) obtained along line 1 and 2 across interphase boundaries marked in the STEM image in Fig.6a. For both concentration profiles, 15 measurements with 1 nm spacing were conducted. In both composition profiles, Cu and Cr phases are visible and minor amounts of Cr and Cu are detected in the Cu and Cr phase, respectively. Furthermore, it can be seen that structural relaxation of the nanostructure has taken place after annealing at 400°C. The grain boundary distortion decreases and distinct boundaries are observed. The grain boundary width is significantly reduced (~ 2 nm) after annealing at 400°C.

## 4. Discussion

### 4.1 Microstructural evolution during HPT deformation and formation of ssss

The hardness and the EDD data indicate that the evolution of the microstructure can be separated into three distinct different regimes: $\varepsilon_{eq}$ smaller than 50, $\varepsilon_{eq}$ between 50 and 150 and $\varepsilon_{eq}$ larger than 200. Hence, the discussion of this part will be subdivided into these three regimes.

#### 4.1.1 $\varepsilon_{eq}$ smaller than 50

In the very early state of the HPT deformation, both Cu and Cr are deformed. Due to the lower strength of Cu usually in some regions a pronounced localized deformation in Cu takes place. But both phases are deformed and the typical formation of dislocation cells, cell blocks and finally an ultrafine grain structure in the Cu and Cr develops, similar as in the pure metals [35]. A saturation of the hardness and refinement for pure Cu and Cr appear at a strain of about 10. There are some small changes in hardness at strains between 10 and 20 associated with the development of a saturation misorientation distribution of the ultrafine grained structure. This phenomena takes place also in the Cu-Cr composites but has not been investigated in detail in this study. The saturation in the structural evolution in the Cu and Cr phase in the composite induces the relatively constant hardness and domain size of Cu and Cr between a strain of 10 – 50. In this region the saturation grain size (domain size) is smaller than the spacing of the Cu and Cr phases.

#### 4.1.2 $\varepsilon_{eq}$ between 50 and 200

At a strain of about 50 the spacing of the phases is about equal or somewhat smaller than the grain size obtained in the pure metals Cu and Cr at this deformation temperature (RT). Between $\varepsilon_{eq} = 50$ and 200 the phase spacing decreases and stabilizes a significant finer grain size in Cu and Cr. This is clearly evident from the decrease of the domain size and the increase in hardness. Due to the deformation of both Cu and Cr the phase spacing decreases continuously till an $\varepsilon_{eq}$ of about 200. Till to this strain the lattice constant of the bcc Cr and fcc Cu and the microstrains do not change significantly. HPT deformation results not only in a significant grain refinement, a high dislocation density and highly distorted non-equilibrium boundaries are reported to occur during SPD [36-39], which might cause the observed slight increase of the microstrain with increasing deformation [39-41].

### 4.1.3 $\varepsilon_{eq}$ larger than 200, the formation of supersaturation

At strains larger than 200, this process of co-deformation seems to break down. The domain size remains constant, only the lattice spacing and microstrains change. This clearly shows the difference in the involved mechanisms. The details of the deformation processes at the grain scale for strains larger than 200 cannot be derived from the present experiments, but the changes of the lattice parameters and strains indicate that on the macro scale the sample are homogeneously deformed i.e. no macro localization takes place. The change of lattice spacing and microstrains indicate a change of the structure on the atomic scale and seems to be a consequence of the observed supersaturation. The hardness reaches a nearly constant value, which increases somewhat until a strain of 500 is reached. Due to the fact, that no further grain size refinement can be observed for strains larger than 200, the contribution of grain size refinement and the formation of ssss on the strength can be roughly estimated which is about 300 and 40-50 $HV_{0.5}$, respectively.

In [42], it is shown that the fragmentation process of W particles in a coarse-grained W-Cu composite deformed by HPT can be described by a fractal distribution. The fractal dimension decrease with decreasing size of the W particles and shows a multi-fractal behaviour at the micro- and nano-levels. The fragmentation behaviour of the Cr particles in the Cu-Cr composite during deformation might be similar as in the W-Cu composite material [42]. It is assumed, that the refinement of the Cr particles continue until a certain size of the particles (domain size of about 10 nm) is reached. After reaching this point, the plastic flow might localize predominantly in the Cu phase. The Cr particles might be steadily flattened, fractured and rebound during continuing deformation similar as in BM or surface SPD [43].

To take a closer look on the change of the lattice spacing, which illustrates the ssss starting point, the relative change of the lattice parameter is plotted in Fig.7 as a function of $\varepsilon_{eq}$. The change of the lattice parameter in the beginning of the deformation ($\varepsilon_{eq}$ <150) is very small, whereas it increases significantly at higher strains at the point the domain size saturates. The increase in the Cr lattice parameter is mainly related to the formation of a Cu ssss in Cr which is confirmed by EDS measurements and APT in [21]. The atomic radius of Cu is 0.1278 nm, the atomic radius of Cr 0.1249 nm [44]. The larger Cu atoms are dissolved in the Cr phase and the interplanar distance and the lattice parameter of Cr increases. Atom probe measurements of HPT processed Cu-Cr composite material in [21] revealed that after 16 rotations ($\varepsilon_{eq}$~250), 7at. % Cu is dissolved in the Cr phase. After applying 25 rotations ($\varepsilon_{eq}$~400), 15at. % Cu in the Cr phase is detected, which is consistent with the relative change of the lattice parameter of Cr. On the contrary, substitution of Cr in Cu should decrease the lattice parameter in Cu. The relative lattice parameter change calculated by the (111, 222) peaks is -0.05, the relative change calculated by the (200, 220, 311) peaks is quite high and positive.

In [21] it was reported, that the formation of ssss in the Cu-Cr system is not symmetric, only somewhat less than 2at. % Cr is dissolved in the Cu phase which is consistent with the EDS data from this work. The bcc Cr phase and the fcc Cu phase have a similar molar volume and the enthalpy of formation of a Cr ssss in fcc Cu is similar to that of a Cu ssss in bcc Cr [15,17]. The energy during BM provides the sufficient $\Delta G_c$ due to the decreasing domain size and the energy stored as crystalline defects to form a Cr ssss in Cu with up to 8wt% Cr in Cu [15, 20]. It might be possible that small amounts of Cr are also dissolved in the Cu phase although they could not be well quantified by APT investigations.

Additional reasons for the change of the lattice parameter can be manifold. In [45], an increase in the lattice parameter in nc materials by various synthesis methods has been observed. It has been related to the existence of rather large strains in nc materials due to dislocations and severely distorted grain boundaries. BM of Cu causes a relative change of the lattice parameter of 0.06%, SPD of Cu (a negative change in the lattice parameter (-0.04%). Ultrafine powder consolidation of Cr enhances the lattice parameter by 0.04%.

SPD leads to generation of a high amount of crystal defects in the material which increases with increasing deformation. An increase of the lattice parameter can therefore be additionally caused by deformation-induced defects like vacancies, see for example [46,47]. Grain boundaries are disordered regions with a certain amount of excess volume in the form of vacancies and vacancy clusters [47]. The excess volume is a function of the grain size and dependant on sample preparation method. During BM which is similar to SPD, the excess volume increases with decreasing grain size [34]. The width of the grain boundary areas in the Cu-Cr composite material is relatively large (~4 nm) in the as-deformed state (Fig.4). The stress field originating from grain boundaries not only induces larger interatomic spacing nearby but also generates lattice strain which increases with increasing deformation. Therefore, the microstrain and lattice parameter have an inherent relationship in nc materials which follows an inverse relation of the grain size, and are strongly dependant on the grain boundary structure and hence, the excess volume [34]. In [48], pure Cu was deformed by SPD and investigated by X-ray diffraction. Although the determined grain size in different crystallographic directions was nearly the same, a significant anisotropy of the microstrain was measured as well. The microstrain in <200> direction was three times larger than <111> direction. A decrease in lattice parameter was observed, which was related to long-range stress field of non-equilibrium grain boundaries with a high density of extrinsic grain boundary dislocations [49].

The lattice parameter and microstrain have an inherent relationship in nc materials [34]. Cu is highly anisotropic and the Young's modulus $E$ varies significantly in different directions ($E_{100}$=66.7 GPa, $E_{110}$=130.3 GPa, $E_{111}$=191.1 GPa) [50]. The measured high microstrain in Cu and the variation of the microstrain in different *(hkl)* directions might be a direct consequence of the different interplanar spacing's between (111) and (200) planes. The unit cell of Cu is contracted in one direction, and expanded in the other. Although Cu atoms are built-in in the Cr lattice, which is accompanied with a steadily increasing lattice parameter, very low microstrain is measured in the Cr phase. In this case, it is assumed that the unit cell increase is isotropic. Additionally, the Young's modulus of Cr (E=279 GPa) is quite high compared to Cu, which can result in a lower microstrain [44].

### 4.1.4. Remarks to the mechanism of the formation of ssss

Due to the positive $\Delta G_c$ and lack of thermodynamic driving forces, a solely diffusion driven mechanism responsible for the formation of ssss can be excluded in the Cu-Cr system [23]. The EDD data reveal, that the formation of ssss during HPT processing starts after reaching the saturation grain (domain) size of Cr and Cu in the composite.

One basic requirement for the beginning of the ssss formation process is therefore the nanometer length scale of the composite structure which is accompanied by a huge portion of interfaces. Considering a mechanism based on pure thermodynamics, a defect-enhanced diffusion driven ssss formation process might be possible due to the high interfacial energy and high number of defects introduced by SPD [23-25]. In principle, the atomic mobility of Cu and Cr is very small at RT [17]. SPD leads to generation of a high amount of crystal defects (vacancies, dislocations, grain boundaries) and a more open crystal structure accelerating diffusion processes under such conditions. For example, the vacancy concentration in SPD pure Cu is very high, which enhances the atomic mobility [47]. In [5], the influence of HPT deformation parameters like the amount of strain, the strain rate and the deformation temperature on the formation of ssss in the Fe-Cu system was investigated. The deformation induced mixing is not strain rate sensitive and is mainly controlled by the accumulated plastic strain. Furthermore, it is only delayed but still occurs if the HPT deformation temperature is reduced to 77K. Therefore, the additional needed enhanced atomic mobility cannot be provided by thermal diffusion as well as vacancies as alternative diffusion media [5].

Our findings furthermore clearly show that the underlying mixing mechanism is controlled by the strain level. The higher the amount of strain, the higher the change of the lattice parameter i.e. more Cu atoms are dissolved in the Cr phase. The process of ssss formation continues even at the highest amounts of applied strain.

Considering the approach proposed in [18,19], multislip shear transfer across heterointerfaces and subsequent interface roughening can induce mixing in Cu-bcc (bcc: Nb, V) nanocomposites if dislocation multiplication and glide mechanisms are geometrically impeded by the nanosized dimensions of the constituent phases. Molecular dynamic simulations investigated shear induced intermixing in non-miscible alloy systems containing fcc and bcc particles [27,28]. For fcc particles (and Fe particles at high strains), dislocation transfer from the matrix into particles is possible. The mixing rate scales with the square of the particle radius. For a diffusive process, a linear dependence would be expected. Nevertheless, distortion of the particle shape and an increase in interface area are the main causes for mixing in the initial stages of mixing. Dislocation transfer across interfaces is not observed in bcc particles (Nb and V), which both keep their original shape.

The formed ssss in the final state are asymmetric, nearly no Cr is dissolved in the Cu phase. Considering a pure mechanical intermixing mechanism as described above, the mixing should be symmetric in the as-deformed state. A possible explanation for the observed asymmetric mixing might be a different decomposition driving force after forced mixing and a different thermal diffusion coefficient [5]. Solute transport can be enhanced by vacancies. Different size distribution of agglomerated vacancy clusters and/or mobility can be found for different SPD metals [51]. In SPD Cu, vacancies are homogeneously distributed, but no literature about vacancies in SPD Cr is yet available. Furthermore, the diffusion coefficient of Cu in bcc Cr at 713 K is $4.1 \times 10^{-17}$ cm²/s, the diffusion coefficient of Cr in fcc Cu is considerable higher ($1.5 \times 10^{-15}$ cm²/s) [52,53]. Rejection of Cr atoms from the Cu phase proceeds more rapidly, i.e. a Cr ssss in Cu is easier to decompose than Cu ssss in Cr.

These findings support the hypothesis that mainly a mechanical intermixing mechanism, which is driven by the total amount of imposed strain are responsible for the formation of ssss in the Cu-Cr system during HPT processing. A "Thermodynamic destabilization" process, as proposed in [5], might account for the observed asymmetric mixing. Finally it should be noted that the possible deformation process of the Cr phase, the steady deformation, fracture, and rebounding of the Cr particles, might induce also an asymmetry in the evolution of ssss, because during the rebounding of Cr particles few Cu atoms remain in the bounding regime of the new formed Cr particle. Such deformation process should result in a strain dependent formation of a ssss more or less independent of the temperature.

**4.2 Microstructural evolution during annealing and hardening by annealing**

In [21], minor grain growth during ageing at 450°C with a final grain size 20–40 nm was observed in the Cu-Cr alloy. Using APT, it was shown that Cr grains are free of Cu after aging and phase separation occurred. The as-deformed state represents a metastable state, which is likely to transfer to a more equilibrium state during annealing at elevated temperatures. Especially grain boundaries, which have a high excess volume, are in a thermodynamic unstable state [34]. Relaxation of the grain boundaries reduces the excess volume, the lattice distortion decreases and long-range strains decrease as well. Grain growth in nc materials can occur at temperatures 0.4 times the melting temperature $T_m$ and even lower [54,55]. The domain size, the lattice parameter and the microstrain is nearly unchanged during annealing at 200°C and 300°C, which clearly indicates that no recovery of the microstructure or decomposition of the formed ssss occurs at this lower annealing temperatures. The recovery behaviour during annealing might be more complex in the case of Cu-Cr alloys due to the formation of ssss during deformation and decomposition during annealing. Before any grain coarsening occurs, a more thermodynamically favoured decomposition and phase separation occur [11]. After annealing at 400°C for 30 min, nearly no microstrain can be measured and the lattice parameter in both phases has reached its original value of the as-deformed state. The dislocation density is reduced and annihilation of defects at grain boundaries and inside the grains occur which is accompanied by relaxation of internal elastic stresses. Diffusion of Cu atoms from the Cr phase is completed and transformation into a Cu-Cr nanocomposite without ssss has taken place which is consistent to previous published data [21]. At the same time, marginal grain growth after annealing at 400°C, which lead to strain free grains has occurred, too. It can be concluded, that decomposition of the formed ssss starts after annealing above 300°C which is completed after annealing for 30 min at 400°C. After the decomposition process is completed, fcc Cu and bcc Cr phases coexist in the composite which causes a delay in subsequent grain coarsening.

In Fig.8, the microhardness at a constant position ($\varepsilon_{eq}$ about 400) in the as-deformed state and after annealing at 200°C, 300°C, 400°C, 550°C and 650°C for 30 min is plotted. Surprisingly, the hardness in the as-deformed state is smaller compared to annealed conditions. The hardness clearly increases if the samples are annealed at 200°C for 30 min. Even at an annealing temperature of 300°C and 400°C, the hardness is still higher compared to the deformed condition. After annealing at 550°C, the hardness reaches the values of the as-deformed sample and decreases slightly after annealing at 650°C. The hardness increase during annealing cannot be explained with a decreasing domain size. In Table 2, the microhardness of a sample deformed for 100 rotations in the undeformed, the as-deformed state ($\varepsilon_{eq}$ about 1600) and after annealing at 100°C, 200°C and 300°C for 30 min is listed. The same annealing hardening behaviour is observed in this sample. After annealing at 100°C, 200°C and 300°C, the hardness of the Cu-Cr sampled deformed for 100 rotations increases from 459 HV to 508 HV. The same hardening behaviour has been observed in SPD FeCu and SPD AgNi nano-composites, which formed ssss during HPT processing, during annealing [56,57].

Various strengthening mechanisms may operate in the Cu-Cr composite in the as-deformed and annealed conditions. In the as-deformed condition, the hardness is mainly governed by the small structural size of the Cu and Cr phase in the composite [58]. Another hardness contribution, which can be added due to the non-equilibrium fraction of Cu in Cr, is a solid solution hardening term, which has also an effect on the Peierls potential [19]. However, only a small solid solution hardening effect in Cu-Cr alloys based on the model of Labusch, in which the modulus and size effect are considered, is determined in [59]. Usually, the strength contribution of solid solution hardening is quite small in this strength regime. High internal stresses indicated by the high amount of microstrain in the Cu phase evolving from the highly distorted interfaces containing a high amount of dislocations in the as-deformed condition, which is typically for SPD processed materials [49], contributes to the hardness as well. Furthermore, the interface is quite diffuse and show a gradual change in composition over a typical width of 4 nm as seen in the APT and TEM investigations [21].

After annealing at 400°C, recovery of the dislocations in the interface and minor coarsening of the structure takes place. Nearly no microstrain is measured in the sample. Additionally, the width of the grain boundaries decreased (~2 nm) compared to the as-deformed state. These findings are consistent to previous published APT data, where the transition length of the Cu-Cr interface is also reduced from ~4 nm to ~2 nm after annealing [21].The formed ssss finally decomposes and a nanocomposite is obtained.

These structural processes are accompanied by an increase of hardness. It can be assumed that in the as-deformed state the high dislocation content in the interfaces enables a relatively easy nucleation of dislocations (effectively the dislocation structure can provide dislocation sources) resulting in a somewhat smaller hardness. With annealing and the ongoing recovery of the dislocations in the interface the dislocation nucleation is hindered and the hardness increases. This phenomenon, hardening by annealing, has been observed before in nc Al produced by SPD [60,61] too. This effect is quite pronounced as it overcomes the loss in strength due to the increasing grain size and the decomposition of ssss (lose of the solid solution hardening contribution). Even in single phase nc materials the prevailing deformation mechanism is not fully understood and still under debate. Hence further investigations, regarding the enhanced hardness at low annealing temperatures and the role of interface structure to the observed strengthening of this complex nanocomposite, are currently conducted.

Even after phase separation at 400°C has occurred, further grain growth is restricted due to the immiscibility of the constituent phases and the observed hardness decrease at higher annealing temperatures might only be due to marginal further grain growth.

## 5. Conclusions

During HPT deformation, the formation of ssss in the Cu-Cr system was obtained. RIETVELD refinement of EDD data has proven to be a powerful tool for particle size and strain broadening analysis of both phases which reveals main features of the deformation process. The findings can be summarized as follows:

(i) At low strains ($\varepsilon_{eq}$<150), microstructural refinement and co-deformation of the Cu and Cr phase take place which reduce the average phase spacing. At higher strains, the minimum particle size of Cr is reached and the plastic flow is mainly localized in the Cu phase. The beginning of ssss formation starts after reaching the minimum particle size of Cr in Cu indicated by a significantly higher relative change of the lattice parameter after this point.

(ii) The formation of ssss during HPT is mainly controlled by a mechanical mixing mechanism and the amount of mixing is controlled by the strain level.

(iii) Structural relaxation of the nanostructure during annealing lead to an enhanced hardness even tough the domain size increases during annealing and the microstrain is reduced to values near zero. The enhanced hardness is attributed to limitations in activating dislocation sources in the annealed material.

**Acknowledgments**

We gratefully acknowledge the financial support by the Austrian Science Fund (FWF): P24141-N20, J3468-N20. We acknowledge the Helmholtz-Zentrum Berlin for provision of synchrotron radiation beamtime at beamline EDDI.

**Figures**

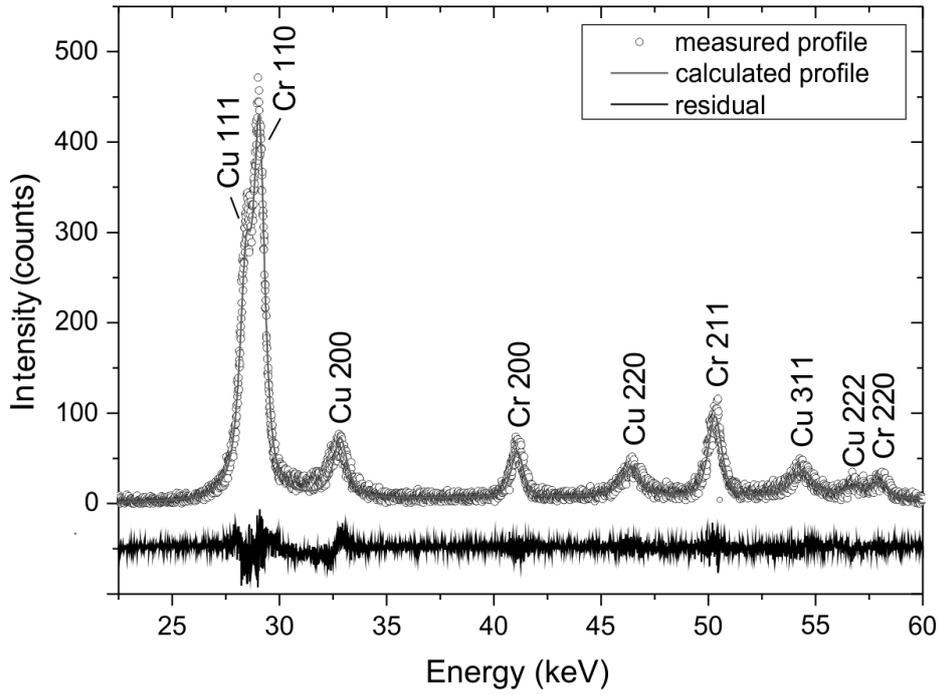

Fig.1 RIETVELD refinement of the EDD pattern of the sample deformed for 25 rotations recorded at r=3.5 mm ($\varepsilon_{eq}$ =400).

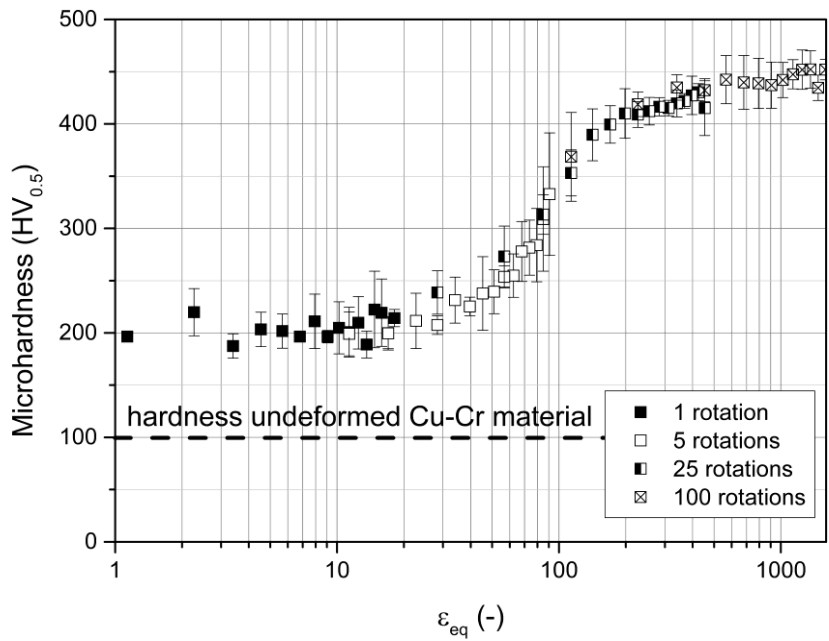

Fig.2 Microhardness of samples deformed for 1, 5, 25 and 100 rotations as a function of the equivalent strain $\varepsilon_{eq}$.

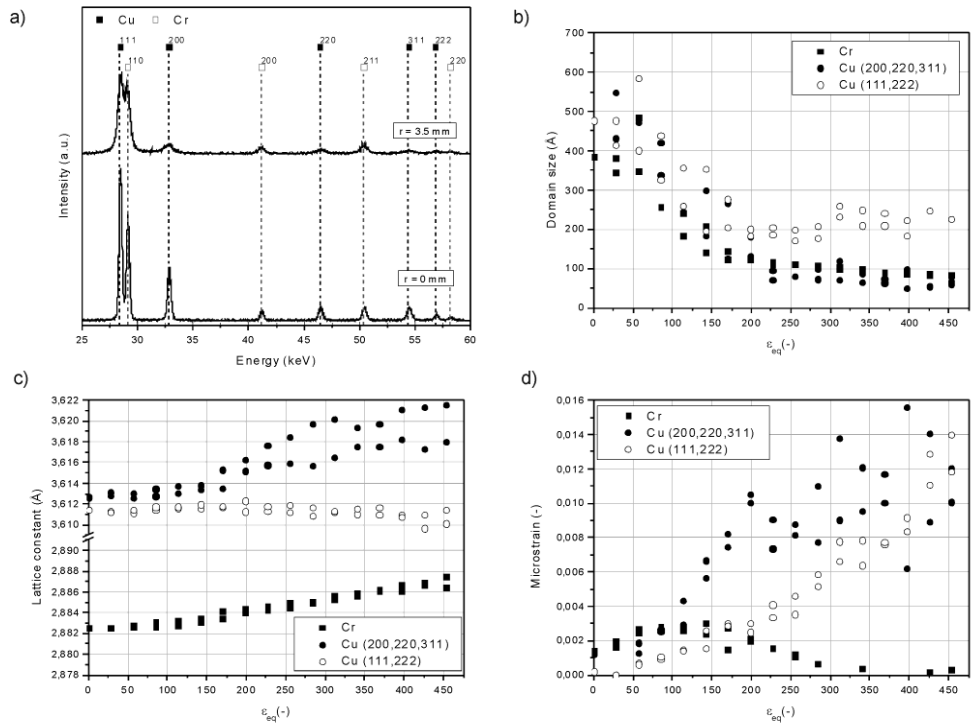

Fig.3 (a) EDD pattern of the sample deformed for 25 rotations recorded at r=0 mm ($\varepsilon_{eq}$~0) and r=3.5 mm ($\varepsilon_{eq}$ =400). (b-d) Calculated values for domain size, lattice parameter and microstrain as a function of $\varepsilon_{eq}$. Values of corresponding positions (positive and negative side of the HPT disk) are plotted, which provides an estimation of the error of the refinement (For example: domain size calculated for r=+1.5mm -1.5 mm, both have a $\varepsilon_{eq}$ of 170).

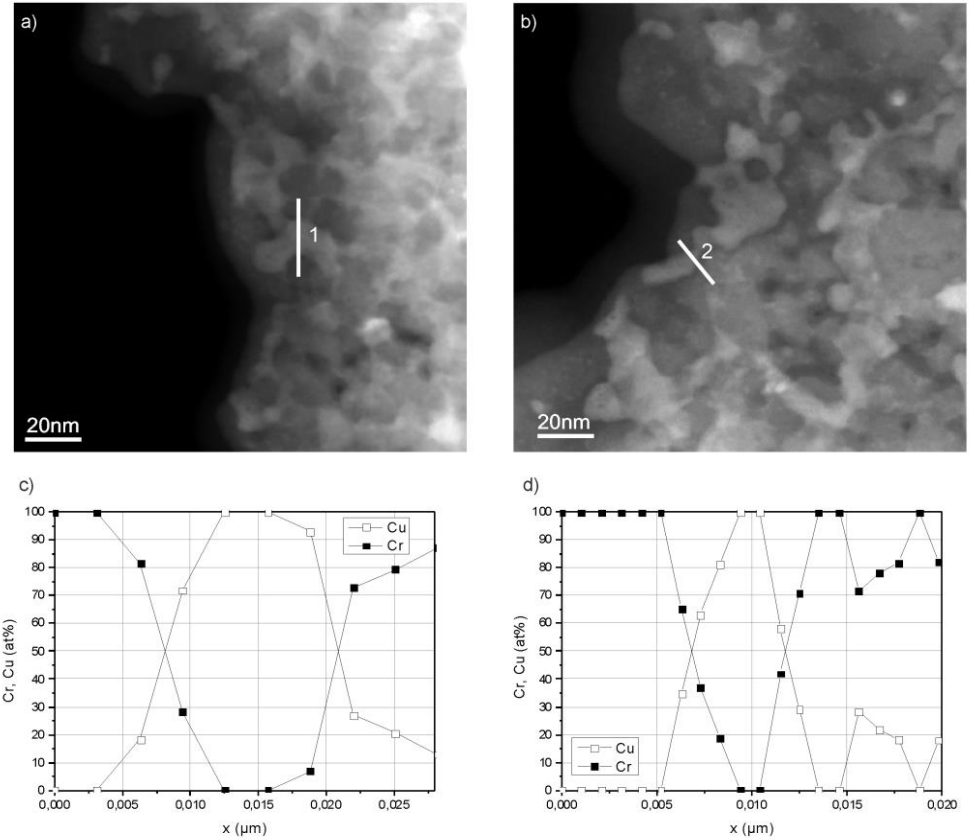

Fig.4 Distribution of Cu and Cr on the nano-scale level: (a) and (b) STEM images of the Cu-Cr sample after HPT processing to $\varepsilon_{eq}$ of 400. (c) Concentrations profiles in atomic percent for Cu and Cr along the line #1 drawn in (a) and (d) concentrations profiles for Cu and Cr along the line #2 drawn in (b).

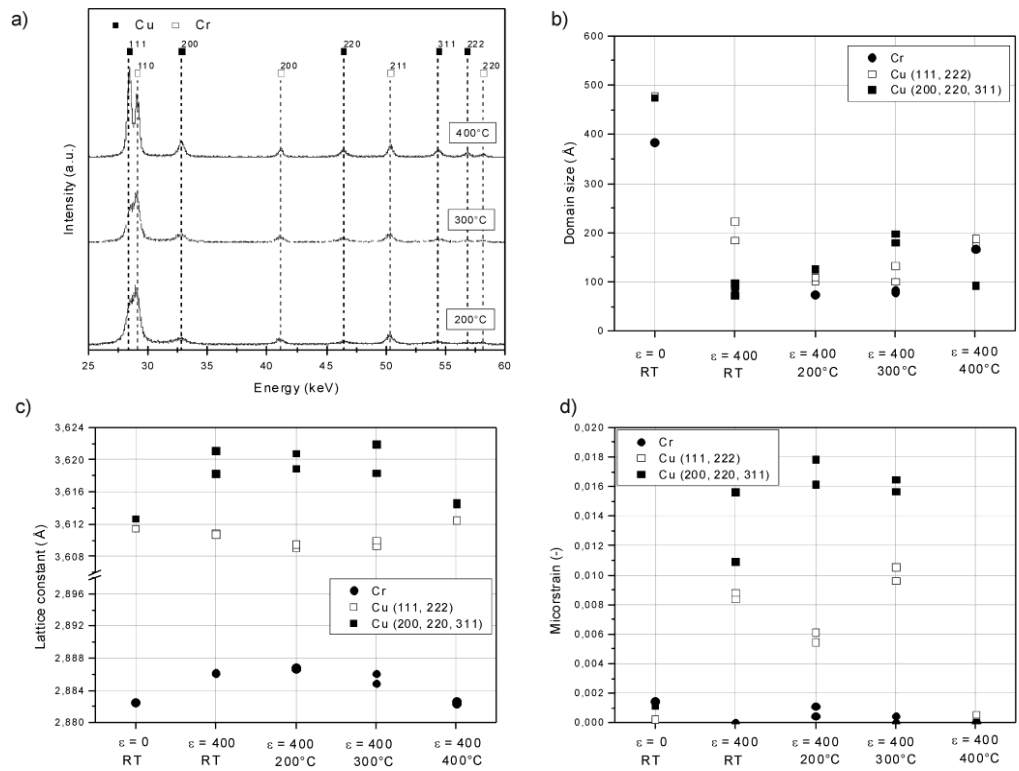

Fig.5 (a) EDD patterns recorded at r=3.5 mm of the sample deformed for 25 rotations ($\varepsilon_{eq}$ =400) after annealing for 30 min at 200°C, 300°C and 400°C. (b-d) Calculated values for domain size, lattice parameter and microstrain for two different deformation conditions ($\varepsilon_{eq}$ ~0 and $\varepsilon_{eq}$ =400) at RT and for three different annealing treatments (200°C, 300°C and 400°C for 30 min) after the deformation ($\varepsilon_{eq}$ =400).

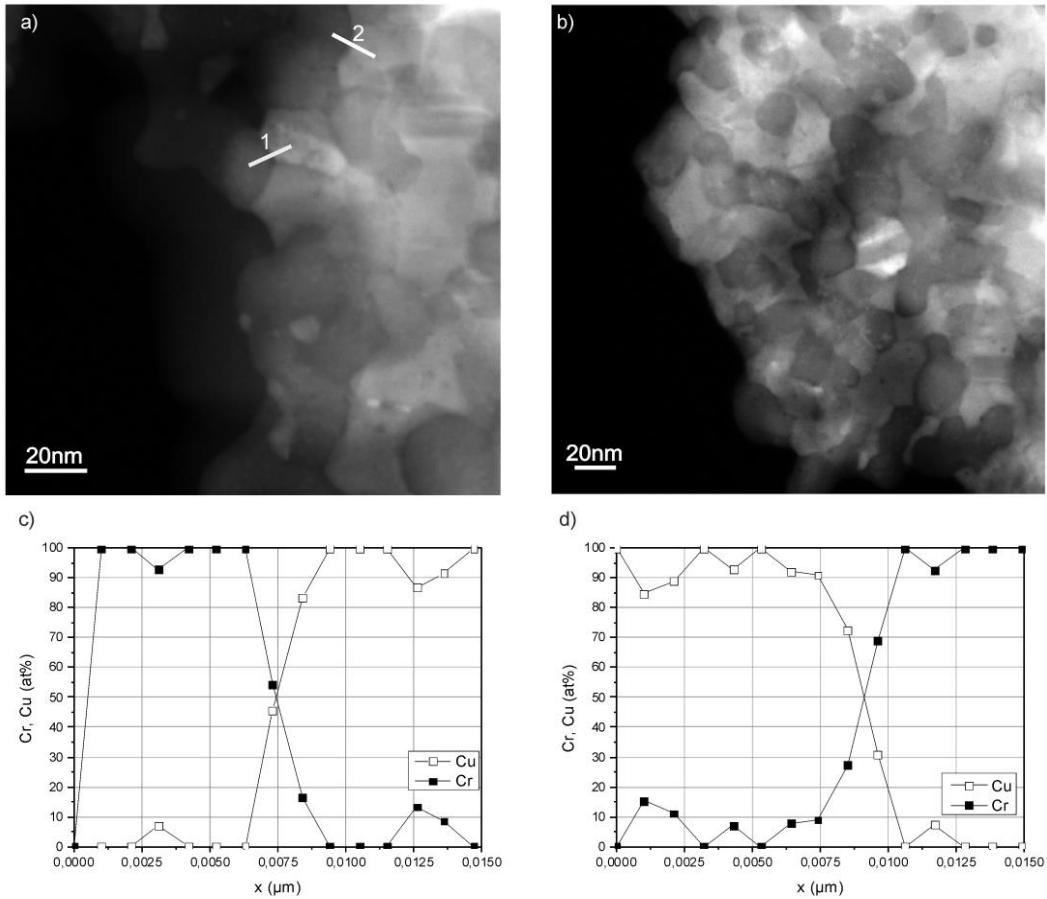

Fig.6 Distribution of Cu and Cr after HPT processing to $\varepsilon_{eq}$ of 400 with additional annealing for 30 min at 400°C: (a) and (b) STEM images of the Cu-Cr sample at different magnifications. (c) Concentrations profiles in atomic percent for Cu and Cr along the line #1 drawn in (a) and (d) concentrations profiles in atomic percent for Cu and Cr along the line #2 drawn in (a).

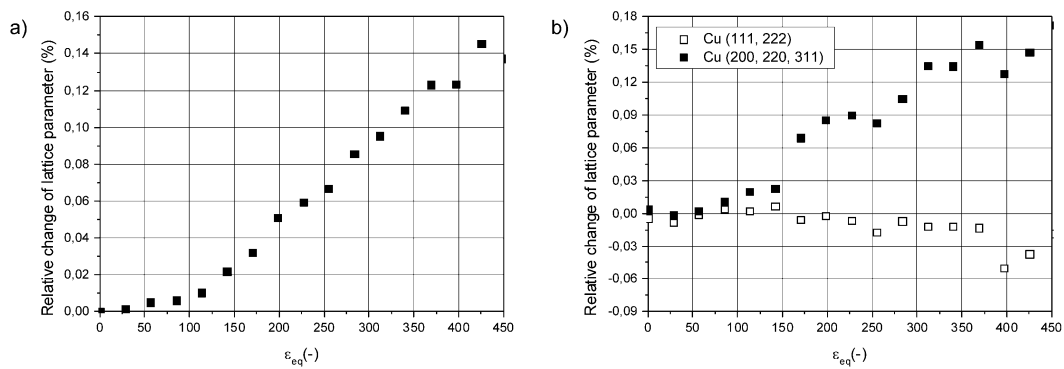

Fig.7 The relative change in lattice parameter as a function of the equivalent strain $\varepsilon_{eq}$ for Cr (a) and Cu (b) calculated by the (111, 222) peaks and (200, 220, 311) peaks, respectively.

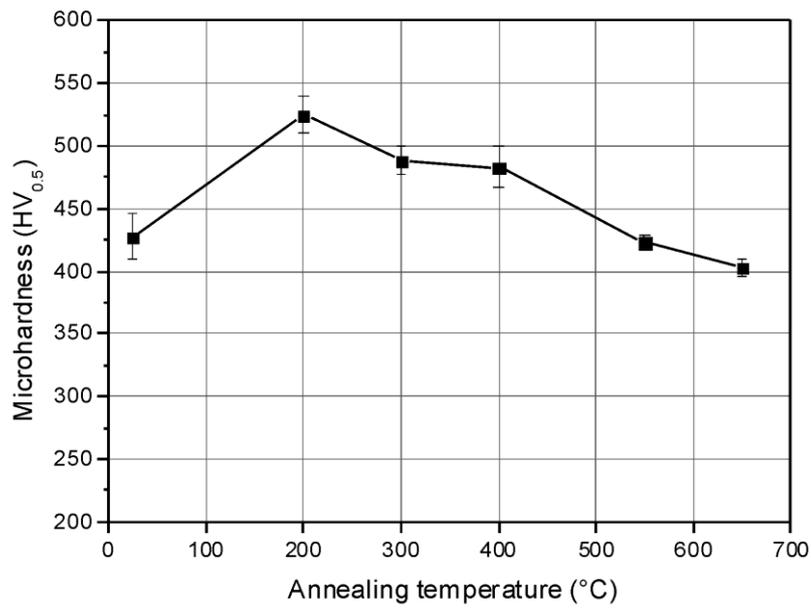

Fig.8 Microhardness at a $\varepsilon_{eq}$ of 400 in the as-deformed condition and after annealing at 200°C, 300°C, 400°C, 550°C and 650°C for 30 min as a function of the annealing temperature.

**Tables**

Table 1: Calculated values for lattice parameter, domain size and microstrain for different deformation conditions ($\varepsilon_{eq}$ of ~0 (near center), 400 and 1590) and various annealing conditions ($\varepsilon_{eq}$ of 400, 200°C-400°C for 30 minutes).

|  |  | $\varepsilon_{eq}$ |  |  |  |  |  |
|---|---|---|---|---|---|---|---|
|  |  | ~0 | 400 | 1590 | 400 (200°C) | 400 (300°C) | 400 (400°C) |
| Cr | lattice parameter (Å) | 2.8825 | 2.8861 ±7.1x10⁻⁶ | 2.8875 | 2.8867 ±1.7x10⁻⁴ | 2.8855 ±8.8x10⁻⁴ | 2.8825 ±2.12x10⁻⁴ |
|  | domain size (Å) | 384 | 89±2.1 | 101 | 75±0.6 | 82±4.1 | 167±1.6 |
|  | microstrain x10⁻³(-) | 1.5 | - | 3.1 | 0.8±0.4 | 0.5 | - |
| Cu (200, 220, 311) | lattice parameter (Å) | 3.6127 | 3.6197 ±2.0x10⁻³ | 3.6189 | 3.6199 ±1.34x10⁻³ | 3.6202 ±1.9x10⁻⁴ | 3.6146 ±2.6x10⁻⁴ |
|  | domain size (Å) | 475 | 73±34.3 | 87 | 126±0.6 | 198±24.4 | 92±2.5 |
|  | microstrain x10⁻³ (-) | 1.2 | 10.9±6.6 | 15.5 | 17.9±2.4 | 16.5±1.1 | - |
| Cu (111,222) | lattice parameter (Å) | 3.6115 | 3.6108 ±1.7x10⁻⁴ | 3.6108 | 3.6091 ±5.6x10⁻⁴ | 3.6094 ±3.8x10⁻⁴ | 3.6124 ±1.8x10⁻⁴ |
|  | domain size (Å) | 477 | 204±26.8 | - | 106±5.0 | 117±22.7 | 181±10.1 |
|  | microstrain x10⁻³ (-) | 0.2 | 8.8±0.5 | 14.6 | 6.2±0.9 | 10.5±1.3 | 0.6±0.1 |

Table 2: Microhardness in the as-deformed condition ($\varepsilon_{eq}$ of 400 and 1590) and after annealing at different temperatures for 30 min.

|  | $\varepsilon_{eq}$ |  |  |  |  |  |  |
|---|---|---|---|---|---|---|---|
|  | ~0 | 400 | 400 (200°C) | 400 (300°C) | 400 (400°C) | 400 (550°C) | 400 (650°C) |
| Microhardness (HV) | 232±45 | 427±18 | 524±15 | 488±11 | 482±16 | 423±6 | 403±7 |

|  | $\varepsilon_{eq}$ |  |  |  |  |
|---|---|---|---|---|---|
|  | ~0 | 1590 | 1590 (100°C) | 1590 (200°C) | 1590 (300°C) |
| Microhardness (HV) | 321±12 | 452±10 | 459±6 | 488±13 | 508±3 |